# Why does hydronium diffuse faster than hydroxide in liquid water?

Mohan Chen,[1] Lixin Zheng,[1] Biswajit Santra,[2] Hsin-Yu Ko,[2] Robert A. DiStasio Jr.,[3] Michael L. Klein,[1,4,5] Roberto Car,[2,*] and Xifan Wu[1,5,*]

[1]*Department of Physics, Temple University, Philadelphia, PA 19122, USA*

[2]*Department of Chemistry, Princeton University, Princeton, NJ 08544, USA*

[3]*Department of Chemistry and Chemical Biology, Cornell University, Ithaca, NY 14853, USA*

[4]*Department of Chemistry, Temple University, Philadelphia, PA 19122, USA*

[5]*Institute for Computational Molecular Science, Temple University, Philadelphia, PA 19122, USA*

**ABSTRACT**

Proton transfer via hydronium and hydroxide ions in water is ubiquitous. It underlies acid-base chemistry, certain enzyme reactions, and even infection by the flu. Despite two-centuries of investigation, the mechanism underlying why hydronium diffuses faster than hydroxide in water is still not well understood. Herein, we employ state of the art Density Functional Theory based molecular dynamics, with corrections for nonlocal van der Waals interactions, and self-interaction in the electronic ground state, to model water and the hydrated water ions. At this level of theory, structural diffusion of hydronium preserves the previously recognized concerted behavior. However, by contrast, proton transfer via hydroxide is dominated by stepwise events, arising from a stabilized hyper-coordination solvation structure that discourages proton transfer. Specifically, the latter exhibits non-planar geometry, which agrees with neutron scattering results. Asymmetry in the temporal correlation of proton transfer enables hydronium to diffuse faster than hydroxide.

The anomalously high mobility of the hydronium, $H_3O^+$(aq), and hydroxide, $OH^-$(aq), ions solvated in water has fascinated scientists since the very beginning of molecular based physical chemistry[1,2]. The two ions can be viewed as opposite topological defects in the fluctuating hydrogen-bond (H-bond) network in liquid water. In this picture, both ions bind to three water molecules *via* donating or accepting H-bonds. Diffusion is not dominated by hydrodynamics, but by a structural process usually referred to as the Grotthuss mechanism[3], in which a proton is transferred from a hydronium to a neighboring water molecule or from a water molecule to a neighboring hydroxide. In this process, a covalent O-H bond breaks while another forms as the topological defect jumps to an adjacent site in the network. Not surprisingly, PT has been intensively investigated, both experimentally and theoretically, for almost a century since the early molecular models[4].

Although the Grotthuss mechanism correctly identifies the origin of fast diffusion, some issues remain unresolved. Experimentally, the diffusivity is obtained via the Nernst equation from the measured electrical conductivity of the ions. While the diffusivity describes the combined effect of hydrodynamic and structural processes, the jump frequency of the protons in the structural process can be extracted from NMR relaxation times. Conductivity[5,6,7] and NMR experiments[8] indicate that hydronium diffuses roughly twice as fast as hydroxide. Predicting the transfer dynamics is difficult as it depends on the cleavage and formation of covalent bonds in a fluctuating liquid medium. Major progress in modeling PT came with the advent of *ab initio* molecular dynamics (AIMD)[9]. In this approach, the forces on the nuclei are derived from the instantaneous ground state of the electrons within density functional theory (DFT)[10,11], while the electrons adjust on the fly and can thereby access bond breaking and forming events. Importantly, the first AIMD study of hydronium and hydroxide in bulk water[12] showed that classical thermal fluctuations easily induce PT events on the picosecond time scale.

In the case of hydronium, the first molecular simulations confirmed the long-held view that transfer involves interconversion of two defect complexes, i.e., the solvated $H_3O^+$(aq) or Eigen ion[13,14], and the solvated $H_5O_2^+$ or Zundel ion[15,16,17,18,19,20]. Even more notable were the results for hydroxide, $OH^-$(aq), which appeared to alternate between two configurations: an unexpected hyper-coordinated form with four acceptor H-bonds and a nearly tetrahedral form with three acceptor H-bonds and, occasionally, a weak donor H-bond. PT only occurred in the latter configuration, suggesting a "presolvation" mechanism, i.e., access to the tetrahedral configuration, was necessary for PT in $OH^-$(aq). Neutron diffraction[21] and core-level spectroscopy data[22] were consistent with the hyper-coordinated structure, indirectly supporting the presolvation picture. However, analysis of neutron scattering data[21] suggested the solvation structure of $OH^-$(aq) had a *pot-like* shape, differing from the *planar* structure predicted by the early AIMD simulations, which employed the generalized gradient approximation (GGA)[23] and treated nuclear dynamics classically. Subsequent path integral AIMD simulations, which treated the nuclei quantum mechanically, refined the model by stressing the fluxional character of the defect complexes, but did not change the basic picture as tunneling was not found to be important[23,24]. In the latter scenario, PT events occur randomly due to thermal and/or quantal fluctuations. However, recent AIMD simulations added a new twist to the story: PT events are highly correlated and happen in bursts consisting of multiple jumps closely spaced in time followed by periods of inactivity[25,26].

The hyper-coordinated hydroxide form revealed by previous simulations breaks the mirror symmetry of the topological defect model between hydronium and hydroxide. Yet, it is not evident why hydronium diffuses faster than hydroxide.

Previous simulations have adopted different flavors of the GGA for the exchange-correlation functional[23,27]. However, such GGA not only overestimates the molecular polarizability and H-bond strength in liquid water, but also tends to grossly underestimate the equilibrium density of the liquid[28,29,30] by neglecting long-range van der Waals (vdW) or dispersion interactions. More significantly, while the predicted diffusivities of $H_3O^+$(aq) are relatively stable from different GGA functionals, those for $OH^-$(aq) can vary by more than one order of magnitude[4,23,24,31] (see Table 1). This large discrepancy inevitably hinders a proper comparison of diffusivities between the two ions.

Here we report AIMD simulations that adopt the hybrid functional PBE0[32,33] and include long-range vdW interactions using a self-consistent implementation[34] of the Tkatchenko-Scheffler (TS)[35] scheme. The resulting PBE0-TS functional is less affected by the spurious self-interaction and better accounts for the molecular polarizability of water, greatly improving the overall description of neat water[36]. Our new data confirm the current picture of hydronium diffusion, namely PTs are highly correlated and occur with relatively high frequency. The effects of the functional approximation are much more pronounced in $OH^-$(aq), i.e., hyper-coordination increases, diffusivity decreases and the effect is accompanied by a strong suppression of multiple jumps. As single jumps become dominant, one may revert back to the picture of uncorrelated random jumps for a rough first order approximation of hydroxide diffusion. We explain this behavior as a consequence of the strongly amphiphilic character through a novel electronic structure analysis, making the lone pair side of OH$^-$(aq) more strongly hydrophilic and its H side more strongly hydrophobic[37]. The diffusion constants of the two water ions and their ratio are in reasonable agreement with experiment.

**RESULTS AND DISCUSSION**

**Proton transfer via the hydronium ion**

The electronic structure of H$_3$O$^+$(aq) is comprised of three bonding electron pairs and one lone electron pair, which are represented by the maximally localized Wannier functions[38,39] in Fig. 1(a). The protons of hydronium are positive and ready to be donated to neighboring water molecules while the oxygen is likely to accept an H-bond from its neighboring water molecules due to the negative lone electron pair. Since the H-bond is mainly attributed to an electrostatic attraction, the ability of donating (accepting) H-bonds can be conveniently measured by the distance separating the negative electrons from the positive nucleus, roughly estimating how positive (negative) the local environment is for a specific proton (oxygen). The resulting distance between electron pairs with respect to the nuclei, as obtained by the PBE0-TS trajectory, are shown in Fig. 1(b) for solvated ions and neat liquid water. Compared to liquid water, the proton of hydronium has a stronger ability to donate an H-bond, while the oxygen of hydronium has a weaker ability to receive one (see Sec. 2 in SI). Therefore, in the absence of PTs, the solvated hydronium is amphiphilic in nature with its proton (oxygen) site being hydrophilic (hydrophobic)[40,41,42]. Hence, H$_3$O$^+$(aq) forms the Zundel or Eigen complexes by stably donating three H-bonds to its neighboring water molecules as shown in Fig. 1(a).

One proton of the hydronium can be transferred to a neighboring water molecule, which in turn is converted to a new ion[43,16]. Moreover, PTs are highly correlated in time evidenced by the preferred bursts of PT events to single PT events.[26] In Figs. 2(a)-(c), we report the frequencies of PTs categorized by the number (single, double, triple, and quadruple) of transfer events during one burst. In general, the PTs obtained from the three AIMD trajectories (PBE, PBE-TS, and PBE0-TS) are all dominated by concerted events with largely preferred double jump events. By analyzing the PBE0-TS trajectory, we illustrate the free energy map in Fig. 3(a) with the length of water wire being a function of the PT coordinate. The analysis confirms the recent discovery that double PTs are associated with the collective compression of a water wire[26]. This concerted behavior enables the proton to diffuse rapidly through two or more water molecules within a single burst, which is enhanced when nuclear quantum effects are considered[44,45]. However, the more physically modeled H-bond network by the vdW interactions and exact exchange has non-negligible effects on the water wire compression and concerted PTs.

The vdW interaction, is an important effect causing denser water than ice under ambient conditions[29,30]. As in the case of pure water[36], the structure of the solution with ions is softened under the influence of vdW interactions. The increased population of water molecules in the interstitial region weakens the H-bond network, while leaving the strength of the short-range directional H-bonds unchanged. As expected, water wires in the softer liquid structure described by PBE-TS can be compressed with a slightly shorter compression length of 0.487 Å compared to that of 0.502 Å in the PBE trajectory (see Sec. 3 in SI). The facilitated water wire compressions encourage more concerted PTs as shown in Fig. 2(b). Consequently, the diffusivity increases as compared to the PBE trajectory (see Table 1).

Electrons, as appropriately described by quantum mechanics, cannot interact with themselves. Yet, all conventional DFT functionals inherit self-interaction error, which artificially overestimates the H-bond strength.[36] This self-interaction error is mitigated by including fractional exact exchange in our PBE0-TS trajectories. Compared to the vdW interactions, the exact exchange directly improves the overestimated H-bond strengths, which also affects the compression of water wire. The H-bond strengths among neighboring water molecules become weaker resulting in a less polarizable liquid towards the experimental direction[36]. The reduced polarizability is mainly provided by the less negative electric environment of oxygen lone pair electrons[36]. The H-bonds binding hydronium to its three neighboring water molecules are also weakened, as evidenced by the shorter distances between protons of the ion and their bonding electron pairs (0.542 Å in PBE0-TS compared to 0.550 Å in PBE-TS trajectories) yielding a less positive electric environment for protons (see Sec. 2 in SI). As a result, the water wire compression with weaker H-bonding becomes less easy as evidenced by the longer compression length of 0.562 Å compared to 0.487 Å in the PBE-TS trajectory (see Sec. 3 in SI). Consistently, slower hydronium diffusion in water is observed compared to that in the PBE-TS trajectory in Table 1.

**Proton transfer via the hydroxide ion**

The $OH^-$(aq) ion is also amphiphilic[37] as determined by its electronic ground state in Fig. 1(b). Based on the same criterion, we can conveniently determine the hydrophobicity and hydrophilicity of $OH^-$(aq) in Fig. 1(b). The oxygen site of hydroxide is hydrophilic and more electronegative than the oxygen site of water, whereas the proton site of hydroxide is hydrophobic and less electropositive compared to the proton site in water (see Sec. 2 in SI). Although both $OH^-$(aq) and $H_3O^+$(aq) are amphiphilic, their electronic origins are different. The hydrophilicity of $H_3O^+$(aq) provided by its protons enables it to donate three H-bonds in the absence of PT. In contrast, the hydrophilicity of $OH^-$(aq) provided by lone-pair electrons enables it to accept either three or four H-bonds, and both

three- and hyper-coordination solvation structures normally occur in the aqueous solution of $OH^-$(aq). The three-coordination solvation structure is tetrahedral-like and encourages PTs via the *presolvation* mechanism[20]; while hyper-coordination strongly disfavors PTs. Therefore, the mirror symmetry of the PT mechanisms between the two ions is broken, and the PT via hydroxide cannot be simply considered as the reverse process of the PT via hydronium by replacing the proton with a "proton hole"[23].

Interestingly, the PTs described by PBE0-TS not only become less frequent but also prefer single jumps to concerted ones, as reported in Fig. 2(f). This mechanism is opposite to the traditional view[26] based on GGA functionals, namely that PTs via $OH^-$(aq) should follow a similar trend as $H_3O^+$(aq), shown in Fig. 1(d). While the PTs become more stepwise, the diffusivity of $OH^-$(aq) also decreases relative to that of $H_3O^+$(aq), approaching a ratio that quantitatively agrees with the experimental value in Table I.

The revised PT mechanism for $OH^-$(aq) implies drastic changes brought by the vdW interactions and the hybrid functional, rather than perturbed water wires compression observed in the $H_3O^+$(aq) solution. Indeed, changes in the solvation structure of $OH^-$(aq) in Fig. 4(d) suggest substantially stabilized hyper-coordination configurations. The PBE functional overestimates the polarizability, yielding over-structured water, and this over-strengthened tetrahedral H-bond network energetically favors the tetrahedral-like three-coordination, i.e., the presolvated structure of $OH^-$(aq). Fig. 4(d) shows that PBE predicts 51% three-coordination and 49% hyper-coordination. With the vdW interactions considered, the H-bond structure of liquid water is softened and facilitates the stabilization of hyper-coordination of $OH^-$(aq). As a result, the percentage of hyper-coordination increases from 49% in the PBE trajectory to 65% in the PBE-TS trajectory. The hyper-coordination is further stabilized to 84% in the PBE0-TS trajectory. The additional amount of hyper-coordination (~19%) is attributed to two physical effects. As far as the H-bond network of the liquid solution is concerned, the exact exchange yields a weakened directional H-bond strength, and generates a further softened liquid water structure, which again helps to stabilize the hyper-coordination structure. In the above, the weakened directional H-bond strength is mainly provided by a less negative environment of the lone pair electrons of liquid water, which is reduced by 2.9% (as measured by the distance between maximally localized Wannier centers and oxygen in Fig 1). At short-range scale, the H-bonding between $OH^-$(aq) and the neighboring waters is much less affected by the exact exchange than that of liquid water. The negative environment due to the lone pair electrons of the hydroxide is only reduced about 1.1%. As a result, the amphiphilic propensity of the solvated hydroxide is promoted, which enables the hydroxide to attract more water molecules further favoring the hyper-coordination structure.

Conventional AIMD theories based on the GGA functionals repeatedly predicted a planar-like solvation structure of hyper-coordinated $OH^-$(aq), i.e., the four hydrogen bonded water molecules accepted by $OH^-$ roughly stay within a plane. We confirm in Fig. 4(a) that the distribution of water molecules surrounding $OH^-$ is relatively flat from the PBE trajectory. The planar structure can be clearly demonstrated by the *planarity* (defined as the distance from one water to the plane formed by the other three water molecules) analysis shown in Fig. 4(e), where the distribution of planarity centers at around zero indicating the dominant planar structure of hyper-coordination. However, the experimental evidence based on the neutron scattering data[21] yields a non-planar solvation pattern.

In the PBE-TS and PBE0-TS trajectories, the H-bond network is modeled more accurately and the hyper-coordination is stabilized. Therefore, more water molecules acquired by hyper-coordination, are attracted to the first coordination shell. These additional water molecules are closer to the oxygen atom of OH$^-$, i.e., a strongly hydrophilic site, and filling in the space close by. Consequently, the hyper-coordination in the PBE0-TS trajectory exhibits a non-planar structure with its *planarity* distribution centered significantly away from zero in Fig. 4(e). Furthermore, the surrounding water molecule density has a *pot-like* structure, shown in Fig. 4(c), as found in the neutron scattering experiments[21]. The agreement strongly suggests that an accurate H-bond description, which has been achieved via PBE0-TS, is crucial to understand the PT mechanism of OH$^-$(aq).

With more stabilized hyper-coordination structures in OH$^-$(aq) from the PBE0-TS trajectory, the presolvated structure (three-coordination) of hydroxide becomes relatively rare. Therefore, the largely suppressed PTs in Fig. 2(f) are expected. However, it is intriguing that the majority of the suppressed PTs are of concerted types, while the frequency of single jump events is marginally influenced. The feature cannot be understood by the presolvation mechanism alone without considering the water wire compression. In this context, it is useful to compare the free-energy landscapes in Figs. 3(a) and (b) for the water wire compression as a function of the concerted (double) PTs coordinate. Consistently, it is found that the energy barrier for a double PT to occur by the water wire compression in OH$^-$(aq) is about 1.9 $k_BT$ larger than the similar energy barrier in H$_3$O$^+$(aq) (see Sec. 3 in SI). The significantly suppressed concerted PTs can be attributed to the energetically stabilized hyper-coordination of OH$^-$. Fig. 4(f) illustrates the changes of coordination number of OH$^-$(aq) with respect to the time before ($t < 0$) and after PTs ($t > 0$). Consistent with the presolvation mechanism, all simulations show OH$^-$(aq) relaxes from three-coordination back to the hyper-coordination after each PT event and vice versa. However, the more energetically stabilized hyper-coordinated OH⁻ in the PBE0-TS trajectory enables a much faster relaxation than that obtained from the PBE and PBE-TS trajectories. On average, the timescale of such relaxation in PBE0-TS trajectory is about 0.3 ps shorter than that of the typical water wire compression (~0.5 ps)[25]. The observed fast relaxation back to the hyper-coordinated OH⁻(aq) is a key to hinder the concerted PT.

**Conclusions**

The origin of the different diffusion mechanisms of the hydrated water ions resides in their electronic ground states. Hence, an accurate theory of the solvent H-bond network is crucial. By utilizing state-of-the-art *ab initio* molecular dynamics, we confirmed PTs via the $H_3O^+$(aq) are frequent, with mostly concerted jumps. By contrast, PT via the solvated hydroxide ion is dominated by a stepwise mechanism governed by the formation of (and rapid relaxation to) a stable non-planar and hyper-coordinated solvation structure. This unique solvation shell, which is structurally consistent with neutron scattering experiments, actively discourages proton transfer in aqueous hydroxide solutions. Since the Stokes diffusions of these two water ions are roughly the same at the level of PBE0-TS theory, which are $(0.76 \pm 0.22) \times 10^{-9}$ and $(0.66 \pm 0.08) \times 10^{-9} m^2/s$ for $H_3O^+$(aq) and OH⁻(aq), respectively, their differences in the nature of concerted PTs against simple PT provide a rational explanation as to why hydronium diffuses faster than hydroxide in water. The different roles played by concerted PT dynamics in $H_3O^+$(aq) and OH⁻(aq) have direct bearing on the interpretation of the NMR experiments, which mostly assumed so far a simple Markovian process to extract the PT rates[8,5]. Nuclear quantum effects (NQEs) play an important role in the dynamics of these two water ions; the concerted PTs will be further enhanced by the delocalized protons[45]. Our main conclusion is expected to be intact since previous studies suggested NQEs do not qualitatively affect the energetics of the solvation structures of the water ions. In this context, the extra stabilization of the hyper-coordination structure of OH$^-$(aq) suggests a likely explanation for the large reported difference in the isotope effect on the transfer rates of the two aqua ions[8], as the deeper free energy well associated to OH$^-$(aq) in our simulation should translate in a comparatively larger quantum zero-point motion effect in OH$^-$(aq) than in H$_3$O$^+$(aq).

**Methods**

We used the Quantum ESPRESSO[46] software to perform simulations based on the density functional theory. The pure water system is composed of 128 $H_2O$. The hydronium system consists of 63 $H_2O$ with one excess proton (127 H atoms and 63 O atoms while the hydroxide system consists of 63 $H_2O$ with one hydroxide ion (127 H atoms and 64 O atoms). In order to reproduce the experimental density of liquid water at ambient conditions, the cubic cells used for ion and pure water simulations have the cell lengths being 12.4 and 15.7 Å, respectively. Only the gamma point was used to sample the Brillouin zone of the supercell. The periodic boundary conditions were utilized with the energy cutoff of plane wave basis being 72 Ry. The Troullier-Martins[47] norm conserving pseudopotentials were employed.

We performed the Car-Parrinello molecular dynamics[9] with the standard Verlet algorithm to propagate nuclear and electronic degrees of freedom. We used a fictitious electronic mass of 150 a.u. to ensure the adiabatic separation between the nuclear and electronic degrees of freedom, and the mass pre-conditioning with a kinetic energy cutoff of 6 Ry was applied to all Fourier components of electronic wave functions.[48] All simulations were performed in the NVT ensemble at 330 K.[49] The ionic temperature was controlled using the Nosé-Hoover chain thermostats[50] with one Nosé-Hoover chain per atom and four thermostats in each chain. The time step was set to be 3.5 a.u. (~0.08 fs). The nuclear mass of deuterium (2.0135 amu) was set for each hydrogen atom in order to accelerate the convergence, while the nuclear mass of oxygen was set to 15.9995 amu. We generated 28, 45, and 32 ps trajectories for the hydronium systems using the Perdew-Burke-Ernzerhof (PBE)[51], PBE with the van der Waals interactions in the form of Tkatchenko and Scheffler[35] (PBE-TS), and PBE-TS with a mixing of 25 percent exact exchange[32,33] (PBE0-TS) functionals, respectively; we also generated 54, 55, and 38 ps for the hydroxide systems using the PBE, PBE-TS, and PBE0-TS functionals, respectively. For the pure liquid water system, we have trajectories of 14, 14, and 25 ps for PBE, PBE-TS, and PBE0-TS trajectories, respectively. We defined the H-bond within a cutoff of 3.5 Å for O-O distance and an H-O-O angle less than 30°.[52] We also used a cutoff of 1.24 Å for the O-H covalent bond.

**Acknowledgements**

This project was supported by U.S. Department of Energy Scidac under Grant No. DE-SC0008726 and No. DE-SC0008626 and partially supported by DMR under Award DMR-1552287. RAD acknowledges partial support from Cornell University through start-up funding and the Cornell Center for Materials Research (CCMR) with funding from the NSF MRSEC program (DMR-1120296). This research used resources of the Argonne Leadership Computing Facility at Argonne National Laboratory, which is supported by the Office of Science of the U.S. Department of Energy under Contract No. DE-AC02-06CH11357. This research also used resources of the National Energy Research Scientific Computing Center, which is supported by the Office of Science of the U.S. Department of Energy under Contract No. DE-AC02-05CH11231.

**Author contributions**

XW, RC, and MLK designed the project. MC and LZ carried out the simulations. MC and LZ performed the analysis. RAD, BS, and H-YK developed methodologies in Quantum ESPRESSO.

XW, RC, MLK, and RAD wrote the manuscript. All authors contributed to the discussions and revisions of the manuscript.

**Additional information**

Supplementary information is available. Correspondence and requests for materials should be addressed to RC and XW.

**Competing financial interests**

The authors declare no competing financial interests.

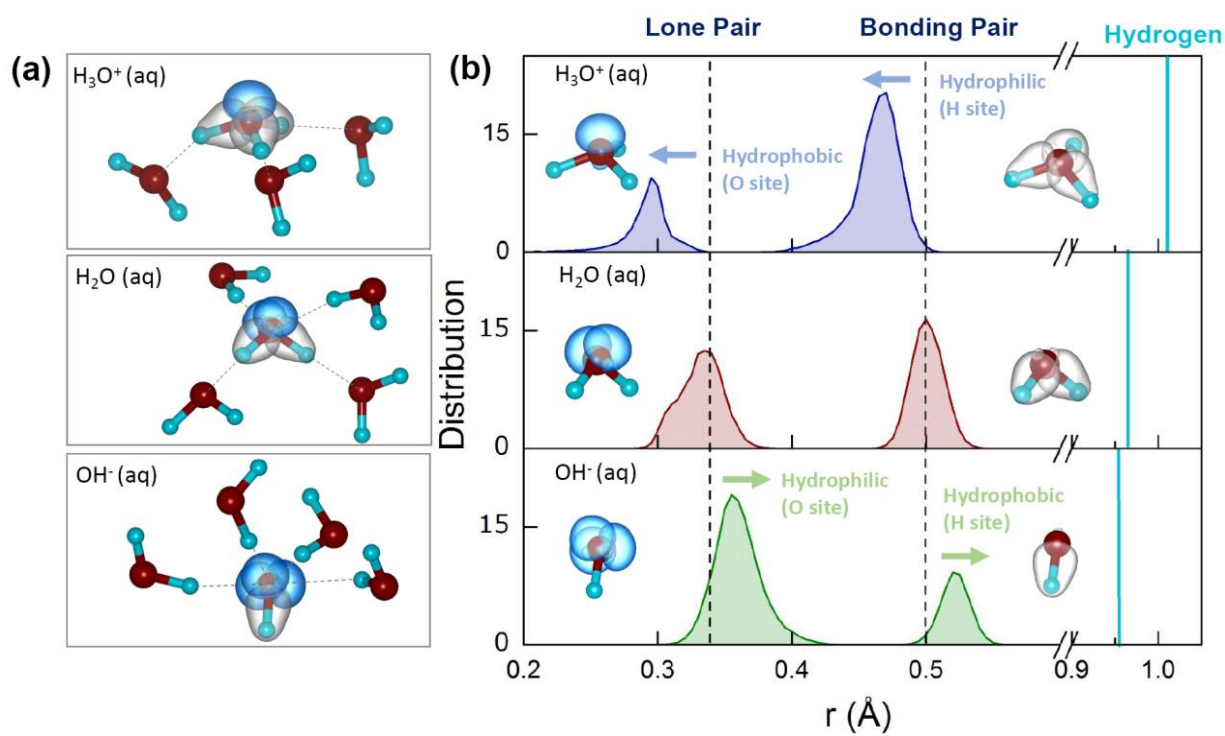

**Figure 1: Electronic structure of the solvated water molecule and water ions from PBE0-TS trajectories.** (a) From top to bottom: solvation structures and maximally localized Wannier functions of $H_3O^+$(aq), $H_2O$(aq), and $OH^-$(aq). $H_3O^+$(aq) donates three H-bonds, $H_2O$(aq) accepts and, respectively, donates two H-bonds, $OH^-$(aq) accepts four H-bonds. Density isosurfaces of maximally localized Wannier functions for lone and bonding pair electrons are depicted in blue and gray, respectively. (b) From top to bottom: distributions of the distances from the intramolecular oxygen of the maximally localized Wannier centers for $H_3O^+$(aq), $H_2O$(aq), and $OH^-$(aq). In each panel, the vertical cyan line indicates the average length of the covalent O-H bond, while the vertical dashed black line indicates the average distance from the intramolecular oxygen of the Wannier centers of lone and bonding pairs. The amphiphilic character of the ions emerges from the comparison with the neutral molecule: an ionic site, oxygen or hydrogen, is hydrophobic (hydrophilic) when the separation between the lone pair and oxygen, or between the bonding pair and hydrogen, is shorter (larger) than the corresponding distance in the neutral molecule.

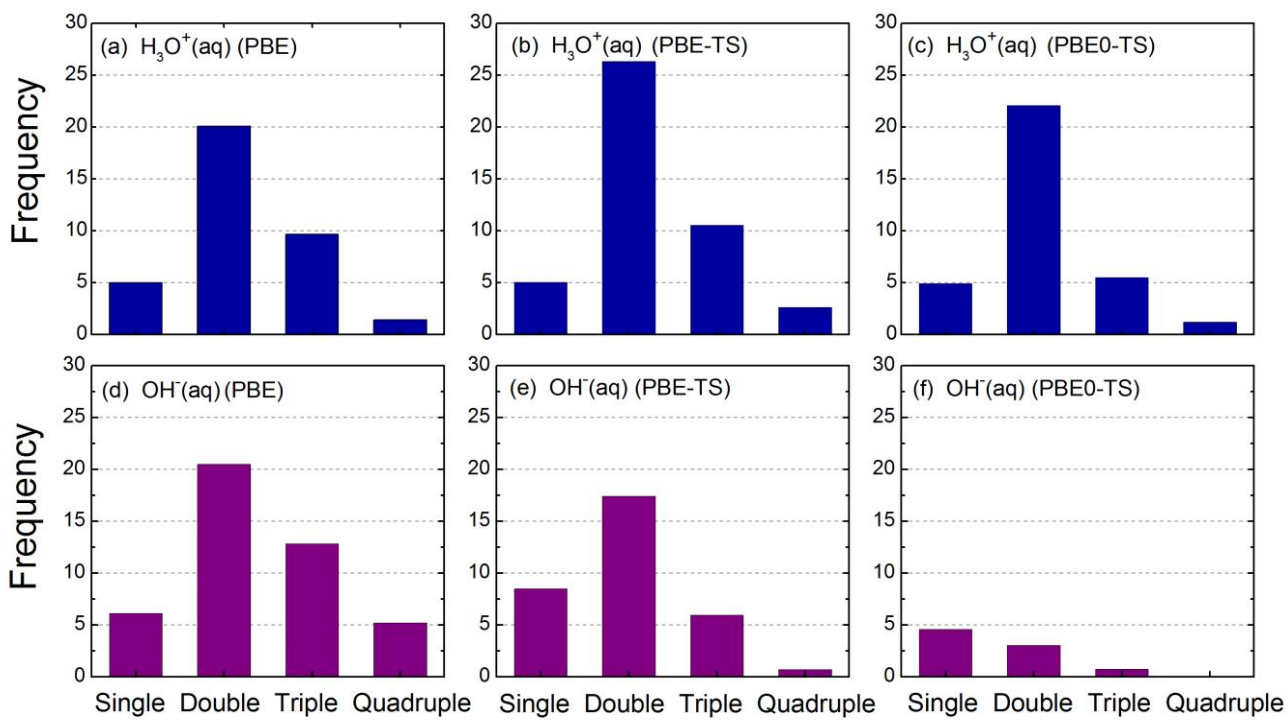

**Figure 2: Frequency of proton transfer (PT) events with three exchange-correlation functionals (PBE, PBE-TS, and PBE0-TS).** The blue bars show the frequency of single, double, triple, and quadruple PTs for $H_3O^+$(aq). The purple bars show the frequency of single, double, triple, and quadruple PTs for $OH^-$(aq). The frequency is calculated by counting the average number of PTs of each kind during a time span of 10 ps. Consecutive jumps separated in time by 0.5 ps or less contribute to multiple, i.e. concerted, PT events. A time lapse of 0.5 ps is the typical observed time scale of compression of a water wire. Events in which a proton returns to its original site within 0.5 ps are considered to be rattling fluctuations and are not included in these counts.

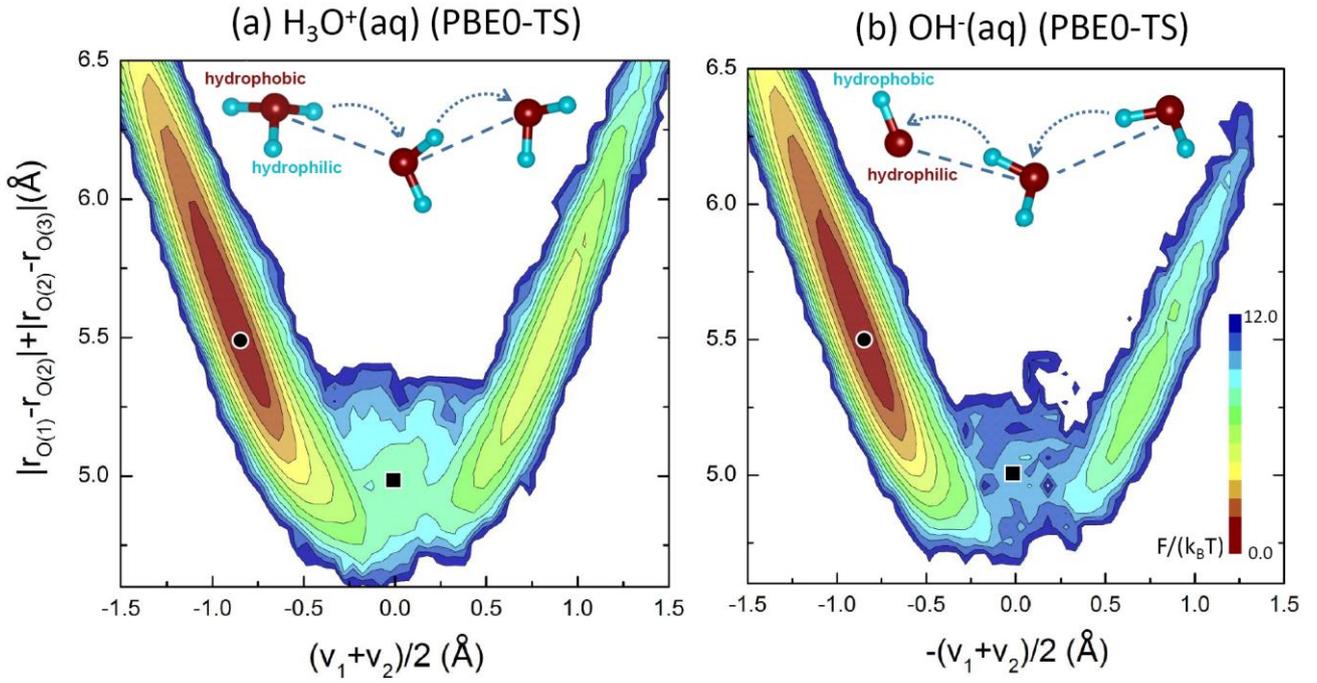

**Figure 3: Free energy maps for water wire compression and double proton jumps with the PBE0-TS functional.** The topographic map on the left is for $H_3O^+$(aq), the one on the right is for $OH^-$(aq). The map gives the isocontours of the free energy in the plane of two collective coordinates, $|r_{O(1)}-r_{O(2)}|+|r_{O(2)}-r_{O(3)}|$, describing the compression of a wire made by three neighboring molecules, and $(v_1+v_2)/2$ or $-(v_1+v_2)/2$, describing the displacement of the two protons attempting a jump, as depicted schematically on top of each map. From left to right, the three oxygens have coordinates $r_{O(1)}$, $r_{O(2)}$, and $r_{O(3)}$, respectively, while the two protons have coordinates $r_{H(1)}$ and $r_{H(2)}$, respectively. The transfer coordinate is $v_1=|r_{O(1)}-r_{H(1)}|-|r_{H(1)}-r_{O(2)}|$ for the first proton, and $v_2=|r_{O(2)}-r_{H(2)}|-|r_{H(2)}-r_{O(3)}|$ for the second proton. Successful double jumps correspond to the configurations on the right of each map, i.e., $v_1+v_2>0$ for $H_3O^+$(aq) or $-(v_1+v_2)>0$ for $OH^-$(aq). All the configurations in a trajectory of a three-water wire, i.e., a triple of bonded molecules, are reported on the left of each map, but only a fraction of these configurations leads to a successful double jump. The free energy gives the relative probability of occurrence of the configurations in each map. The most stable configurations (left of the maps) are indicated by a black dot and the corresponding free energy is set equal to zero. The saddle points for double PT are indicated by the black squares. The corresponding free energy barriers are 7.5 and 9.4 $k_BT$ for $H_3O^+$(aq) and $OH^-$(aq), respectively. More details on barrier calculations are reported in Sec. 3 of SI.

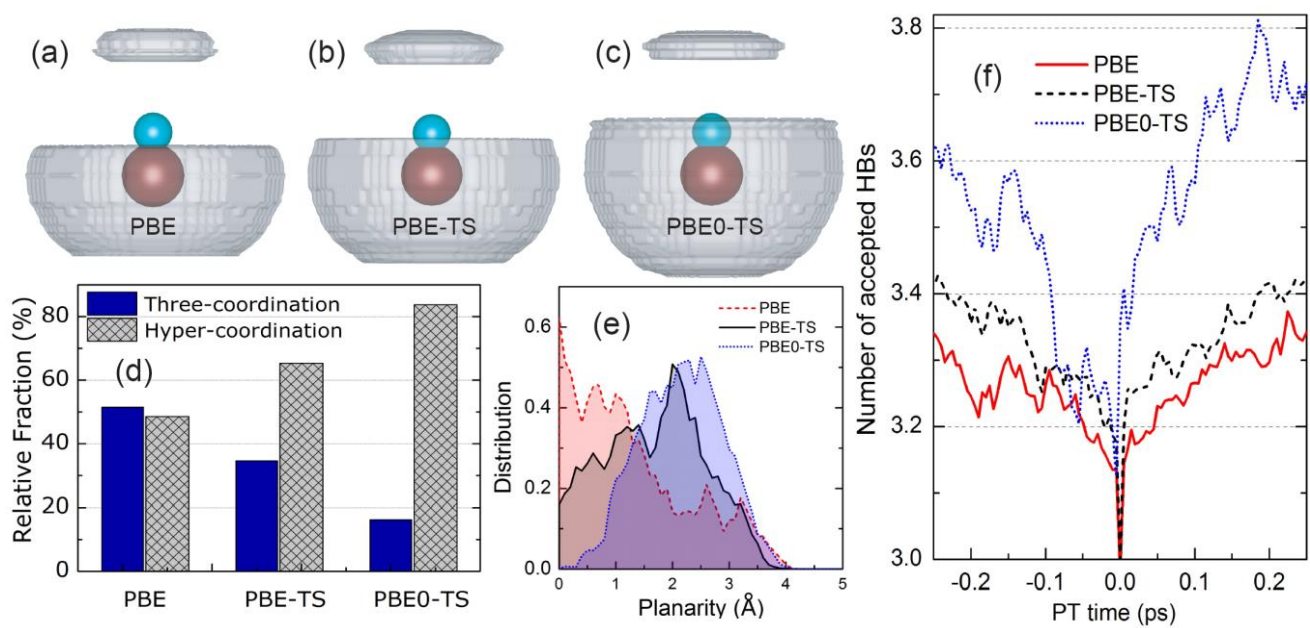

**Figure 4**: **Solvation structures of OH⁻(aq) with three functional approximations (PBE, PBE-TS, and PBE0-TS).** (a), (b) and (c) Isosurfaces representing the spatial distribution of the oxygen sites of the solvating molecules in the hyper-coordinated structure of OH⁻(aq) with three functionals. The hydroxide ion has the hydrogen (cyan sphere) pointing upward and the oxygen (red sphere) pointing downward. (d) Relative fraction of three- and hyper-coordinated solvation structures with three functionals. In the three-coordinated structure OH⁻(aq) accepts three H-bonds, in the hyper-coordinated structure it accepts four H-bonds or more. (e) Planarity distribution of the hyper-coordinated structures with three functionals. The planarity order parameter is defined by the distance between a coordinating oxygen atom and the plane formed by three other coordinating oxygen atoms. (f) Coordination number (number of acceptor H-bonds) of OH⁻(aq) before ($t<0$) and after ($t>0$) a PT event. PT events are very fast (~0.005 ps) on the time scale of the plot. Thus OH⁻(aq) is always unambiguously defined and we can follow its evolution by adopting the Lagrangian point of view.

**Table 1. Computed ratios from the diffusion coefficients $D$ of $H_3O^+$(aq) ($D^+$) and $OH^-$(aq) ($D^-$), the experimental diffusivity data are computed from the limiting molar conductivities $\lambda$ of $H_3O^+$(aq) ($\lambda^+$) and $OH^-$(aq) ($\lambda^-$) measured at 28[5,7] and 25 °C[6]**

| $D$ ($10^{-9}$ m$^2$/s) | $D^+$ | $D^-$ | $D^+/D^-$ |
|---|---|---|---|
| **PW91** | 3.24 | 18.5 | 0.18 |
| **BLYP** | 2.83 | 1.92 | 1.47 |
| **HCTH/120** | 3.25 | 0.44 | 7.39 |
| **PBE** | 10.8±2.7 | 18.2±3.7 | 0.59±0.27 |
| **PBE-TS** | 12.8±1.9 | 8.3±1.6 | 1.54±0.53 |
| **PBE0-TS** | 8.3±1.9 | 3.7±0.4 | 2.24±0.75 |
| **Exp. (H$_2$O)** | 9.6[5], 9.4[7] | 5.4[5], 5.2[7] | 1.77[5], 1.80[7] |
| **Exp. (D$_2$O)** | 6.9[5], 6.7[7] | 3.2[5], 3.1[7] | 2.15[5], 2.15[7] |
| $\lambda$ ($\Omega^{-1}$ cm$^2$ mol$^{-1}$) | $\lambda^+$ | $\lambda^-$ | $\lambda^+/\lambda^-$ |
| **Exp. (H$_2$O)** | 364.0[5], 351[7] | 206.0[5], 195[7] | 1.77[5], 1.80[7] |
| **Exp. (D$_2$O)** | 261.6[5], 252[7] | 121.5[5], 117[7] | 2.15[5], 2.15[7] |
| $\lambda$(H$_2$O)/$\lambda$(D$_2$O) | 1.39[5], 1.39[7] 1.364[6] | 1.70[5], 1.67[7] | |

The simulation data for six exchange-correlation functionals, i.e., PW91[53], BLYP[54,55], HCTH/120[56], PBE, PBE-TS, and PBE0-TS, are reported. The data for the first three functionals are from Ref. [23], those for last three functionals are from the present simulation. We also show the standard deviations for computed $D$. The deuterium mass was used for both ions and water molecules in all AIMD simulations listed, while the experimental data (Exp.) based on both H and D are listed. See Sec. 4 of SI for more information on the procedures to compute diffusivities based on the AIMD simulations.

The experimental diffusivity data were computed based on the Nernst equation $D = \frac{RT}{F^2}\lambda$, where $R$ is the gas constant, $T$ is the temperature, and $F$ is Faraday's constant.

# Supplementary Materials for

## Why does hydronium diffuse faster than hydroxide in liquid water?

Mohan Chen, Lixin Zheng, Biswajit Santra, Hsin-Yu Ko, Robert A. DiStasio Jr.,

Michael L. Klein, Roberto Car[*] and Xifan Wu[*]

*Corresponding author. Email: **rcar@princeton.edu**, **xifanwu@temple.edu**

**The file includes**

Figures S1-S2

Tables S1-S2

References

## 1. Amphiphilicity of solvated hydronium and hydroxide in water

In section "results and discussion" of the main paper, we discussed the amphiphilic nature of both solvated hydronium $H_3O^+$(aq) and hydroxide $OH^-$(aq) ions in liquid water as obtained from the PBE0-TS trajectories. Besides the electronic origins explained in the main paper, the amphiphilicity of both water ions can be further supported by the solvation structures as represented by the radial distribution functions (RDFs) $g(r)$ and their covalent and hydrogen (H) bonding properties, which are reported in Fig. S1 and Table S1, respectively.

We first draw our attention on the amphiphilicity of the solvated hydronium ion. On one hand, the hydrophilicity of $H_3O^+$(aq) originates from its protons, which are more electropositive than those in pure water as already explained via maximally localized Wannier functions[39] in the main paper. As a result, the protons of $H_3O^+$(aq) form stronger H-bonds with the surrounding water molecules than those in neat water. The stronger H-bonds are confirmed by a closer and narrower first coordination shell in the $g_{H^*O^w}(r)$ of $H_3O^+$(aq) than that in the $g_{H^wO^w}(r)$ of liquid water, as shown in Fig. S1 (a) and (b), respectively. In the above, $H^*$, $O^w$, and $H^w$ denote the hydrogen of $H_3O^+$(aq), oxygen of $H_2O$(aq), and hydrogen of $H_2O$(aq), respectively. Due to the hydrophilic nature of proton sites, the covalent bond length of $H_3O^+$(aq) (1.01 Å) is found to be longer than that of neat water (0.97 Å) as listed in Table S1. Consistently, the number of donor H-bonds of $H_3O^+$(aq) averaged over the trajectory is found to be 1.0 per proton in Table S1, which indicates that each proton of $H_3O^+$(aq) is robustly donated to one neighboring water molecule in the absence of proton transfer. In contrast, the average number of donor H-bonds in pure water is only found to be 0.87 per proton, implying a relatively more fragile H-bond in liquid water under thermal fluctuations. On the other hand, the hydrophobicity of the $H_3O^+$(aq) is attributed to the less electronegative lone electron pairs of its oxygen atom. In this regard, the H-bonding between oxygen of $H_3O^+$(aq) and the protons of surrounding water molecules is found to be weaker than its counterpart in neat water. This can be further evidenced by a more separated and broader first coordination shell in $g_{O^*H^w}(r)$ of $H_3O^+$(aq) than that in $g_{H^wO^w}(r)$ of neat water, as illustrated in Fig. S1 (d) and (e), respectively. Here $O^*$ denotes the oxygen atom of $H_3O^+$(aq). Consistently, the average number of acceptor H-bonds in $H_3O^+$(aq) is 0.35 per lone electron pair, which is significantly smaller than the number of 0.87 in neat water.

By the same token, we next discuss the auxiliary evidence supporting the amphiphilicity of the solvated hydroxide ion. Different from $H_3O^+$(aq), the hydrophilicity of $OH^-$(aq) originates from its more electronegative lone electron pairs of oxygen atom than those in neat water. Therefore, the H-bonds formed between the oxygen of $OH^-$(aq) and its neighboring water molecules are stronger than those in pure liquid water. The stronger H-bonds can be supported by the following two aspects. First, the water molecules in the first coordination shell of $OH^-$(aq) are getting closer compared to those in liquid water, which can be seen from the comparison between the $g_{O^*H^w}(r)$ of $OH^-$(aq) and the $g_{O^wH^w}(r)$ of $H_2O$(aq) as shown in Fig. S1 (f) and (e), respectively. Second, the hydrophilic oxygen site (three lone pairs) allows each lone electron pair to accept 1.31 H-bonds on average, which is greater than the value of 0.87 in neat water (two lone pairs) as listed in Table S1. On the contrary, the hydrophobicity of the $OH^-$(aq) arises from the less electropositive environment of its proton site. On average, the number of donor H-bonds of the solvated hydroxide is 0.47 as reported in Table S1, suggesting a more weakened H-bond when compared to the number of 0.87 per proton in pure liquid water. The less easily donated proton of hydroxide is also consistent with the observed shorter covalent bond length of 0.96 Å in $OH^-$(aq) compared to the length of 0.97 Å in water in Table

S1. Due to the weakened H-bonding strength, the first peak of $g_{H^*O^w}(r)$ of OH⁻(aq) shifts outwards and presents a broader distribution when compared to that of $g_{H^wO^w}(r)$ of neat water in Fig. S1 (c) and (b), respectively.

## 2. Water wire compression lengths and barriers

In section "proton transfer via hydronium ion" of the main paper, the water wire compression length, $L_c$, is defined as the difference of length between a resting water wire, $L_r$, in the absence of proton transfer (PT) and a compressed water wire, $L_p$, during the concerted (double) PTs events. All water wires considered are 3-water wires, each one of which is composed of triply bonded molecules. In the above, the $L_r$ is determined by the average length of water wire corresponding to the most stable configuration with the lowest free energy, which is denoted by the black dot in Fig. 3. By extending the criterion adopted by Tuckerman et al. in describing single PT events,[23] the concerted (double) PTs are considered to be occurring while the double PTs coordinate $(v_1+v_2)/2$ is within the range of [-0.1 Å, 0.1 Å]. Based on the interval, the length of the compressed water wire, $L_p$, can be computed, which is located at the saddle point denoted by the black square in Fig. 3. The computed lengths of $L_r, L_p$ and the resulting $L_c$ of $H_3O^+$ (aq) are shown in Table S2 for all three density functionals of PBE, PBE-TS, and PBE0-TS, respectively.

The free energies of $E_r$ and $E_p$ can be respectively determined by the abovementioned two stages, i.e., the most stable configuration corresponding to a rest water wire and the compressed water wire during the concerted (double) PTs events. Furthermore, the energy barrier for the concerted (double) PTs is defined as the difference of free energy between the resting water wire and compressed water wire as $\Delta E = E_r - E_p$. The resulting $\Delta E$ for both OH⁻(aq) and H₃O⁺(aq) are 7.5 and 9.4 $k_B T$ for PBE0-TS.

## 3. Diffusivities of hydronium and hydroxide ions in water

The current simulations do not allow us to quantitatively determine the diffusion coefficients of the solvated ions. A well converged diffusivity from computations requires larger simulation boxes and longer simulation time, which are still computationally challenging considering the expensive cost in evaluating the non-local exact exchange of PBE0-TS functional. Nevertheless, we can robustly describe the qualitative differences in the diffusion coefficients of H₃O⁺(aq) and OH⁻(aq) by analyzing the trajectories predicted by all three functional approximations.

As shown in Fig. 2S (a) and (b), the diffusion coefficients $D$ of both OH⁻(aq) and H₃O⁺(aq) are calculated from the slope of the mean square displacement (MSD) as a function of time $\tau(t)$ according to the Einstein equation $D = \frac{1}{6}\frac{d\tau(t)}{dt}$. To get better statistics, the MSD has been computed over the average of uniformly divided segments of the equilibrated PBE0-TS AIMD trajectories, each segment of which is 12 *ps* long and the displacement between neighboring segments is 3 *ps*. The PTs in OH⁻(aq) and H₃O⁺(aq) are respectively dominated by single and double jumps as we describe in the main text. Therefore, the single and double PTs are closely associated with the compression of H-bonded 2-molecule and 3-molecues water wires, respectively. In this regard, by considering a typical lifetime of H-bond[52] ~ 1.5 *ps* and the characteristic water wire compression time ~ 0.5 *ps*,[25] we consider the diffusions of ions are uncorrelated by the 3 *ps* time interval separating the above segments in a trajectory.

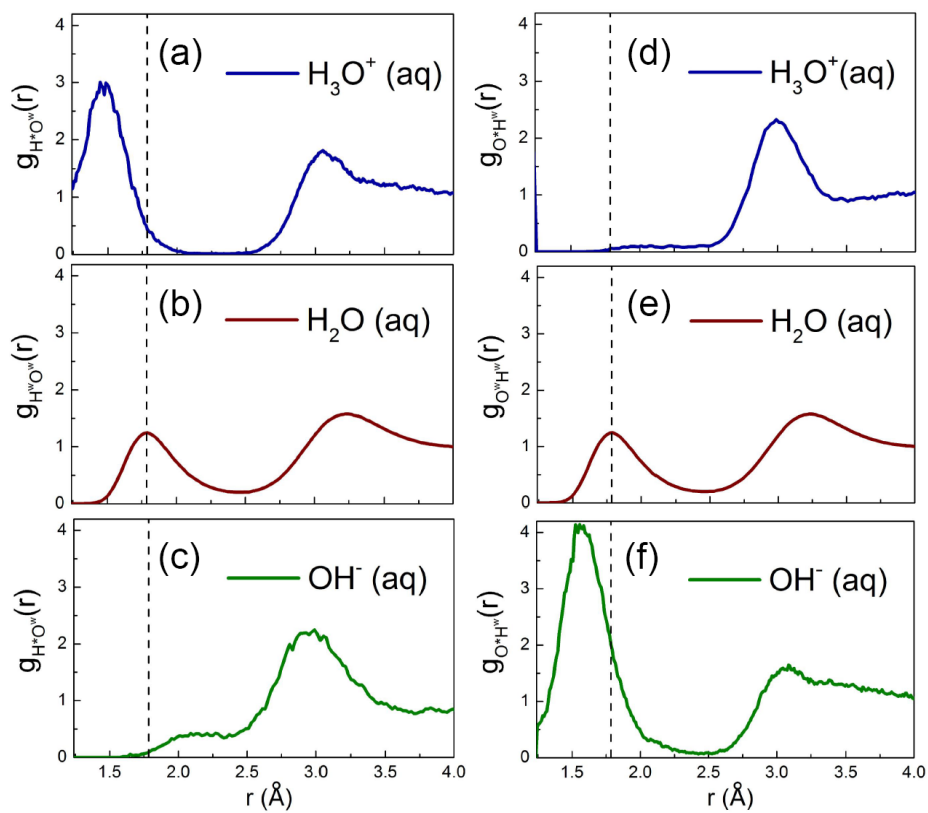

**Figure S1.** Radial distribution functions (a) $g_{H^*O^w}(r)$, (b) $g_{H^wO^w}(r)$, and (c) $g_{H^*O^w}(r)$ for H$_3$O$^+$(aq), H$_2$O(aq), and OH$^-$(aq), respectively. Radial distribution functions (d) $g_{O^*H^w}(r)$, (e) $g_{O^wH^w}(r)$, and (f) $g_{O^*H^w}(r)$ for H$_3$O$^+$(aq), H$_2$O(aq), and OH$^-$(aq), respectively. H$^*$, O$^*$, H$^w$, and O$^w$ refer to hydrogen of ion, oxygen of ion, hydrogen of water, and oxygen of water, respectively.

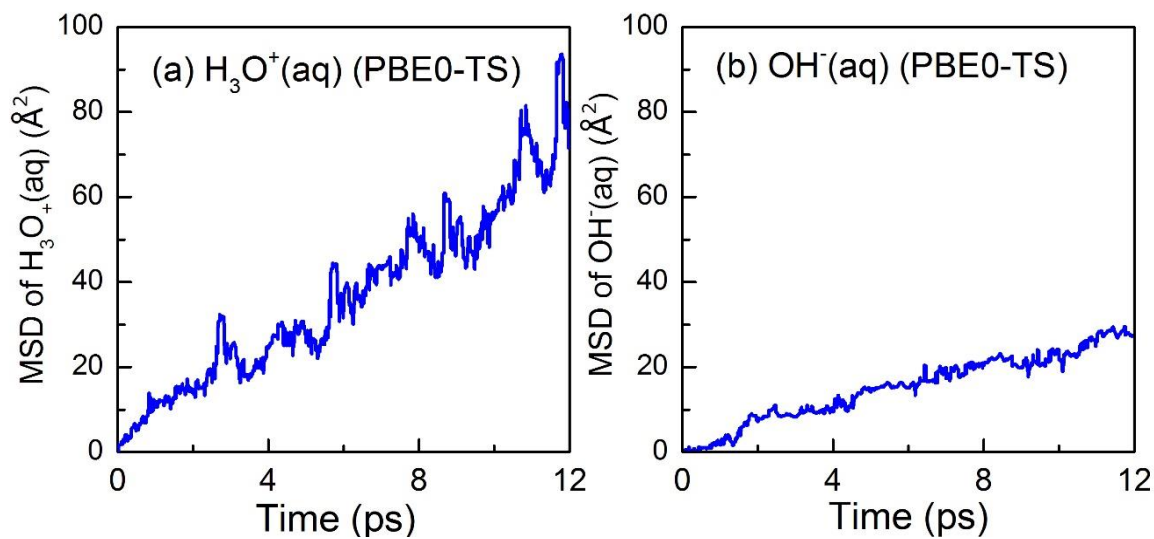

**Figure S2.** Mean square displacement (MSD) of $H_3O^+$(aq) and $OH^-$(aq) computed from the PBE0-TS trajectories as a function of time. The double and single PT events dominate the MSDs of $H_3O^+$(aq) and $OH^-$(aq), respectively. Therefore, the diffusion of $H_3O^+$(aq) is qualitatively twice faster than that of $OH^-$(aq) at ambient conditions as obtained from the PBE0-TS trajectories. The quantitatively computed diffusion coefficients are listed in Table 1 in the main text.

**Table S1**: Covalent and hydrogen bonding properties of $H_3O^+$(aq), $H_2O$(aq), and $OH^-$(aq) obtained by the PBE0-TS exchange-correlation functional. $r_{cov}$ refers to the length of oxygen-hydrogen covalent bond and is defined as the first peak position in the radial distribution function $g_{O^*H^*}$ in ion solutions and the first peak position in $g_{O^w H^w}$ of neat water. We note that $O^*$, $H^*$, $O^w$, and $H^w$ represent the oxygen of ion, hydrogen of ion, oxygen of water molecule, and hydrogen of water molecule, respectively. $H_3O^+$(aq), $H_2O$(aq), and $OH^-$(aq) have one (three), two, and three (one) lone (bonding) electron pairs, respectively.

|  | $r_{cov}$ (Å) | Acceptor H-bonds per lone electron pair | Donator H-bonds per proton |
|---|---|---|---|
| $H_3O^+$(aq) | 1.01 | 0.35 | 1.00 |
| $H_2O$ (aq) | 0.97 | 0.87 | 0.87 |
| $OH^-$ (aq) | 0.96 | 1.31 | 0.47 |

**Table S2**: Based on the free-energy maps in Figure 3 in the main text, we list the length ($L_r$) of a resting water wire, the length ($L_p$) of a compressed water, and the compression length of water wire defined as $L_c=L_r-L_p$ from PBE, PBE-TS, and PBE0-TS functionals. These numbers quantitatively indicate that water wire compression is slightly easier (0.487 Å) using the PBE-TS functional than that from the PBE functional (0.502 Å). However, the water wire compression becomes more difficult (0.562 Å) with the usage of the PBE0-TS functional. The $L_r$ was chosen based on averaging the lengths of water wires located at the lowest free energy point (0.0 $k_BT$) and its nearby area that has free energies smaller than 0.2 $k_BT$.

|         | $L_r$ (Å) | $L_p$ (Å) | $L_c$ (Å) |
|---------|-----------|-----------|-----------|
| **PBE**     | 5.499     | 4.999     | 0.502     |
| **PBE-TS**  | 5.500     | 5.013     | 0.487     |
| **PBE0-TS** | 5.532     | 4.970     | 0.562     |